\documentclass[journal,twoside]{IEEEtran}
\IEEEoverridecommandlockouts
\usepackage{url}
\usepackage{cite}
\usepackage{amsmath,amssymb,amsfonts}
\usepackage{algorithmic}
\usepackage{graphicx}
\usepackage{soul}

\usepackage{textcomp}
\usepackage{xcolor}
\def\BibTeX{{\rm B\kern-.05em{\sc i\kern-.025em b}\kern-.08em
    T\kern-.1667em\lower.7ex\hbox{E}\kern-.125emX}}

\usepackage{balance}
\usepackage[unicode=true, bookmarks=false, breaklinks=false,pdfborder={0 0 1},colorlinks=false]{hyperref}
\hypersetup{colorlinks,citecolor=black,urlcolor=black,linkcolor=black}
\usepackage[linesnumbered,ruled,vlined]{algorithm2e}
\SetKwInput{KwInput}{Input}                
\SetKwInput{KwOutput}{Output}              
\SetKw{KwBy}{by}

\usepackage{multirow}
\usepackage{threeparttable} 

\usepackage{fixmath}
\usepackage{nicefrac}
\usepackage{resizegather}

\DeclareMathOperator*{\argmax}{arg\,max}

\makeatletter
\newcommand{\linebreakand}{%
\end{@IEEEauthorhalign}
\hfill\mbox{}\par
\mbox{}\hfill\begin{@IEEEauthorhalign}
}
\makeatother

\begin{document}

\title{The Smooth Trajectory Estimator for LMB Filters
\thanks{
This research has been funded by the Australian Research Council through the Linkage project LP200301507.}
}

\author{Hoa Van Nguyen\textsuperscript{1}, Tran Thien Dat Nguyen\textsuperscript{1}, Changbeom Shim\textsuperscript{1}, Marzhar Anuar\textsuperscript{2}\\ 
\textsuperscript{1}\textit{School of Electrical Engineering, Curtin University, WA 6102, Australia } \\
\textsuperscript{2}\textit{Department of Agriculture Malaysia, Universiti Putra Malaysia, Serdang 43400, Malaysia } \\
\textsuperscript{1}\{hoa.v.nguyen;t.nguyen1;changbeom.shim\}@curtin.edu.au, \textsuperscript{2}marzhar80@gmail.com
}

\maketitle

\begin{abstract}
This paper proposes a smooth-trajectory estimator for the labelled multi-Bernoulli (LMB) filter by exploiting the special structure of the generalised labelled multi-Bernoulli (GLMB) filter. We devise a simple and intuitive approach to store the best association map when approximating the GLMB random finite set (RFS) to the LMB RFS. In particular, we construct a smooth-trajectory estimator (i.e., an estimator over the entire trajectories of labelled estimates) for the LMB filter based on the history of the best association map and all of the measurements up to the current time. Experimental results under two challenging scenarios demonstrate significant tracking accuracy improvements with negligible additional computational time compared to the conventional LMB filter. The source code is publicly available at \url{https://tinyurl.com/ste-lmb}, aimed at promoting advancements in MOT algorithms.
\end{abstract}

\markboth{PrePrint: IEEE  ICCAIS 2023, Nov. 2023, Hanoi, Vietnam}{PrePrint: IEEE  ICCAIS 2023, Nov. 2023, Hanoi, Vietnam}


\section{Introduction}

Multi-object tracking (MOT) 
aims to identify varying numbers of objects and their trajectories in the presence of noisy data. Because of noisy sensors resulting in misdetections and false alarms, as well as the randomness of object 
disappearances and appearances (i.e., the object's birth and death processes), solving the MOT problems is extremely more challenging than the single-object tracking problem~\cite{blackman1999design,bar2011tracking}. Additionally, MOT plays crucial roles in various applications ranging from aerospace~\cite{reid1979algorithm}, robotics~\cite{mullane2011a,hoa2019jofr}, surveillance~\cite{blackman1999design,bar2011tracking}, to cell biology~\cite{rezatofighi2015multi,dat2021celltracking}. Although there are various approaches to MOT, most align with three principal frameworks: joint probabilistic data association (JPDA)\cite{rezatofighi2015joint,bar2011tracking,musicki2004joint}, multiple hypothesis tracking (MHT)\cite{blackman1999design,bar2011tracking,reid1979algorithm}, and random finite set (RFS)~\cite{mahler2007statistical,mahler2014advances}.

The RFS framework, a recent development in MOT, has gained significant attention in the past twenty years due to its capacity to manage intricate tracking scenarios. This approach considers the multi-object state as a finite set, utilising finite set statistics techniques for temporal estimations. Given its robust mathematical foundation, several RFS-based filters have emerged, including the probability hypothesis density (PHD)\cite{mahler2003multitarget}, cardinalised probability hypothesis density (CPHD)\cite{mahler2007cphd}, multi-Bernoulli (MB)\cite{mahler2007statistical,vo2008cardinality}, and Poisson multi-Bernoulli mixture filter (PMBM)\cite{williams2015marginal}. Notably, these filters estimate only the multi-object states, omitting trajectory details.

For estimating the multi-object trajectories (i.e., the history of multi-object states), one can utilise the labelled RFSs by augmenting unique labels/identities to individual object states~\cite{blackman1999design,mahler2007statistical}.  Significantly, trajectories play a pivotal role in capturing the motion and interaction of objects within a setting. 
Concurrently, labels are essential in differentiating individual trajectories and conveying trajectory-related information. 
In particular, based on the theory of labelled RFSs,  the generalised labelled multi-Bernoulli (GLMB) filter~\cite{vo2013glmb,vo2014glmb} stands as the inaugural exact closed-form solution for multi-object tracking, efficiently approximated using Gibbs sampling~\cite{vo2016efficient,shim2023linear}. Owing to its reliability and adaptability, the GLMB filter has been utilised in diverse applications like lineage tracking~\cite{dat2021celltracking,bryant2018generalized}, track-before-detect~\cite{papi2015generalized,hoa2019tsp}, distributed MOT~\cite{fantacci2018robust,hoa2021distributed}, path planning~\cite{beard2017void,hoa2020aaai,hoa2022multi}, multi-sensor \cite{vo2019multisensor}, multi-scan~\cite{vo2019multiscan}, and large-scale~\cite{beard2020solution} MOT. The LMB filter~\cite{reuter2014lmb}, an approximation of GLMB's first moment, significantly curtails association hypotheses by categorising tracks and measurements into distinct, statistically independent groups. Yet, being a GLMB filter derivative, the LMB filter can encounter issues like \textit{track fragmentation and label switching}.

Smoothing produces superior tracking performance compared to filtering  since smoothing considers  the full  history of the states up to the current time, whereas filtering only considers the most recent state~\cite{doucet2009tutorial,briers2010smoothing}. Another weakness of filtering compared to smoothing in the context of MOT is the \textit{track fragmentation and label switching} since low existence probability tracks might be killed and re-born with new track labels~\cite{vo2019multiscan}.  GLMB smoothing has been proposed in \cite{vo2019multiscan} by solving multi-scan MOT problems via Gibbs sampling, and further generalised to multi-scan multi-sensor GLMB  in~\cite{moratuwage2022multi}. However, computing an exact multi-object posterior in the multi-scan MOT is an NP-hard problem and is typically approximated via Gibbs sampling to solve the multi-dimensional assignment problems. An alternative approach entails a proficient GLMB smoothing algorithm centred on multi-object trajectory estimates rather than multi-object posteriors, as presented in \cite{dat2019glmb}. This method maintains computational complexity akin to the conventional GLMB filter but offers considerable enhancements in tracking performance.

In this work, drawing inspiration from \cite{dat2019glmb}, we introduce a novel smooth trajectory estimator algorithm for LMB, termed \textbf{STE-LMB}. This algorithm performs on multi-object trajectory estimates rather than multi-object posteriors, leveraging the unique architecture of the GLMB filter. Specifically, during the LMB filter's update phase, while converting the GLMB density as an LMB density, we concurrently record the \textit{best} association map linking each labelled track to the measurements at the present time step. This process yields a comprehensive association history for each labelled track, enabling the crafting of a smooth trajectory estimator across each labelled estimate's full trajectory, anchored on all associated measurements up to the present time. Thus, our smooth trajectory estimator executes forward filtering from an object's birth time to the present and then backward smoothing from the current time to its inception for every labelled track. This methodology not only mitigates \textit{track fragmentation and label switching} but also curtails localisation discrepancies typically seen in the conventional LMB filter.

This paper is structured as follows: Section II delivers essential background on labelled RFSs and the LMB/GLMB filters. Our proposed smooth trajectory estimator algorithm is detailed in Section III. Numerical experiments and a comparative analysis with the standard LMB filter are covered in Section IV. Finally, Section V summarises our conclusions.

\section{Background}
The section offers foundational knowledge on labelled RFSs, and an overview of the associated LMB/GLMB filter.

\subsection{Notations} Using the notation from \cite{reuter2014lmb}, lowercase characters like $x, \mathbf{x}$ denote single-object states, whereas uppercase ones like $X, \mathbf{X}$ symbolise multi-object states. Boldface characters such as $\mathbf{X}, \boldsymbol{\pi}$ represent labelled states and their densities, and blackboard characters like $\mathbb{X}, \mathbb{L}$ stand for spaces. For any set $X$, $\mathcal{F}(X)$ signifies the class of its finite subsets. The indicator function of $X$ is represented as $1_{X}(\cdot)$ and its cardinality as $|X|$. For a function $f$, its multi-object exponential is defined as $f^X = \prod_{x \in X} f(x)$, with $f^{\emptyset} = 1$. We also introduce the generalised Kronecker delta function $\delta_Y(X)$, which is one if $X=Y$ and zero otherwise. The inner product $\int f(x) g(x) dx$ is concisely represented as $\langle f,g \rangle$.

\subsection{Labelled Random Finite Sets} 

\noindent \textbf{Labelled multi-object representation.}~ At time $k$, a surviving object is described by a labelled state $\mathbf{x} = (x,\ell)$. Here, the state $x \in \mathbb{X}$, while the distinct label $\ell = (s,\alpha)$ in $\mathbb{L}$ combines its birth time, $s$, and a unique identifier, $\alpha$, distinguishing objects with identical birth times. The trajectory for the shared label $\ell = (s,\alpha) $ from time $s$ to $t$ is the time-sequential sequence $\tau = [(x_s,\ell),\dots,(x_t,\ell)]$. The collection of surviving objects, each with a unique label, at time $k$ is encapsulated by the labelled multi-object state $\mathbf{X} \subseteq \mathbb{X}\times\mathbb{L}$. The set of labels from $\mathbf{X}$ is symbolised as $\mathcal{L}(\mathbf{X}) = \{ \mathcal{L}(\mathbf{X}) : \mathbf{X} \in \mathbf{X} \}$.

Within the window interval $\{j:k\}$ for a sequence of labelled multi-object states $\mathbf{X}_{j:k}$, the trajectory corresponding to the label $\ell \in \bigcup_{i=j}^k \mathcal{L}(\mathbf{X}_i)$ is defined as follows~\cite{vo2019multiscan}:
\begin{align}
    \mathbf{x}^{(\ell)}_{s(\ell):t(\ell)} &= \bigg[(x^{(\ell)}_{s(\ell)},\ell),\dots,(x^{(\ell)}_{t(\ell)},\ell)\bigg],\\
    s(\ell) &= \max(j,\ell[1,0]^T), \label{eq:label_birth_time} \\
    t(\ell) &= s(\ell) + \sum_{i=s(\ell)+1}^{k} 1_{\mathcal{L}(\mathbf{X}_i)}(\ell),  \label{eq:label_end_time}
\end{align}
 where the start and end times of label $\ell$ within the window interval $\{j:k\}$ are denoted by $s(\ell)$ and $t(\ell)$, respectively. As a result, the sequence $\mathbf{X}_{j:k}$ can be represented by
\begin{align}
    \mathbf{X}_{j:k} = \bigg\{ \mathbf{x}^{(\ell)}_{s(\ell):t(\ell)}: \ell \in \bigcup_{i=j}^k \mathcal{L}(\mathbf{X}_i)  \bigg\}. 
\end{align}

\vspace{1mm}
\noindent\textbf{Labelled multi-Bernoulli (LMB) RFS.~} An LMB RFS denoted as $\mathbf{X}$ is characterised by the parameter set $\boldsymbol{\pi} = \{r(\ell),p(\cdot,\ell)\}_{\ell \in \mathbb{L}}$. Here, $r(\ell)$ signifies the existence probability of label $\ell$, whilst $p(\cdot,\ell)$ represents the spatial distribution corresponding to label $\ell$, ensuring that $\int p(x,\ell)dx = 1$. The LMB density is detailed in ~\cite{reuter2014lmb}:
\begin{align} \label{eq_LMB_RFS}
	\boldsymbol{\pi}(\mathbf{X}) = \triangle(\mathbf{X}) w(\mathcal{L}(\mathbf{X}))p^{\mathbf{X}},
\end{align}
where $\triangle(\mathbf{X}) = \delta_{|\mathbf{X}|}(\mathcal{L}(|\mathbf{X})|)$ is a distinct label indicator;  $w(L) = \pi(\emptyset) \prod_{\ell \in L} \big(r(\ell)/(1-r(\ell))\big)$; $\boldsymbol{\pi}(\emptyset)=\prod_{\ell \in \mathbb{L}} (1 - r(\ell))$;  $p(\mathbf{x}) = p(x,\ell)$. For clarity, we represent the LMB density as $\boldsymbol{\pi} =\{r(\ell),p(\cdot,\ell)\}_{\ell \in \mathbb{L}} = \big\{ (w(I),p): I \in \mathcal{F}(\mathbb{L}) \big\}$. 

\vspace{1mm}
\noindent\textbf{$\delta$-Generalised labelled multi-Bernoulli ($\delta$-GLMB) RFS.~}A $\delta$-GLMB RFS delineates the statistical interrelations between objects by accounting for multiple hypotheses. These consist of a set of track labels, denoted as $I \in \mathcal{F}(\mathbb{L})$, and a respective association history symbolised as $\xi \in \Xi$.  The $\delta$-GLMB density is detailed in\cite{vo2016efficient}
\begin{align} \label{eq_GLMB_RFS}
	\boldsymbol{\pi}(\mathbf{X}) = \triangle(\mathbf{X}) \sum_{(I,\xi) \in \mathcal{F}(\mathbb{L} ) \times \Xi} w^{(\xi)}(I) \delta_I(\mathcal{L}(\mathbf{X})) [p^{(\xi)}]^{\mathbf{X}}
\end{align}
For simplicity, we denote the $\delta$-GLMB density as  $\boldsymbol{\pi} = \big\{ (w^{(\xi)}(I),p^{(\xi)}): (\xi,I) \in \Xi \times \mathcal{F}(\mathbb{L}) \big\}$. 

\subsection{Generalised Labelled Multi-Bernoulli (GLMB) Filter}
Given the current $\delta$-GLMB filter density at time $k$ as $\boldsymbol{\pi}$ in \eqref{eq_GLMB_RFS} and the LMB birth model, the $\delta$-GLMB density at time $k+1$ (indicated by $+$) based on measurement set $Z_+$ is defined accordingly~\cite{vo2016efficient}:
\begin{align}
    \boldsymbol{\pi}_+(\cdot|Z_+) = \bigg\{ \big(w^{(\xi,\theta_+)}_{Z_+}(I_+), p_{Z_+}^{(\xi,\theta_+)}\big): (\xi,\theta_+,I_+) \bigg\}
\end{align}
where $\xi \in \Xi$, $\theta_+ \in \Theta_+$, $I_+ \in \mathcal{F}(\mathbb{L}_+)$, and
\begin{align}
    w^{(\xi,\theta_+)}_{Z_+}(I_+) &\propto \sum_{I\subseteq \mathbb{L}} w^{(\xi)}(I)1_{\mathcal{F}(I\uplus \mathbb{B}_+)}(I_+) w_{Z_+}^{(I,\xi,I_+,\theta_+)},\\
w_{Z_+}^{(I,\xi,I_+,\theta_+)} &= 1_{\Theta_+(I_+)}(\theta_+)[1-\bar{P}_S^{(\xi)}]^{I-I_+}[\bar{P}_S^{(\xi)}]^{I \cap  I_+} \notag \\&\times [1-r_{B,+}]^{\mathbb{B}_+ - I_+}r_{B,+}^{\mathbb{B}_+ \cap I_+} [\bar{\psi}_{Z_+}^{(\xi,\theta_+)}]^{I_+},\\
\bar{P}_S^{(\xi)}(\ell) &= \langle p^{(\xi)}(\cdot,\ell),P_{S}(\cdot,\ell) \rangle, \\
\bar{\psi}_{Z_+}^{(\xi,\theta_+)}(\ell_+) &= \langle \bar{p}^{(\xi)}_{+}(\cdot,\ell_+),\psi^{(\theta_+(\ell_+))}_{Z_+}(\cdot,\ell_+)\rangle, \\
\bar{p}^{(\xi)}_{+}(x_+,\ell_+) &= 1_{\mathbb{L}}(\ell_+) \dfrac{ \langle P_S(\cdot,\ell_+)f_+(x_+|\cdot,\ell_+),p^{(\xi)}(\cdot,\ell_+) \rangle }{\bar{P}_S^{(\xi)}(\ell_+)} \notag \\
&\quad\quad+1_{\mathbb{B}_+}(\ell_+)p_{B,+}(x_+,\ell_+), \\
p_{Z_+}^{(\xi,\theta_+)}(x_+,\ell_+) &= \dfrac{\bar{p}^{(\xi)}_{+}(x_+,\ell_+) \psi_{Z_+}^{(\theta_+(\ell_+))}(x_+,\ell_+) }{\bar{\psi}_{Z_+}^{(\xi,\theta_+)}(\ell_+)}.
\end{align}
Here, $\Theta_+$ represents the collection of positive 1-1 mappings $\theta_+ : \mathbb{L}_+ \rightarrow \{0 : |Z_+|\}$. The survival probability of the label $\ell$ is given by $P_S(\cdot,\ell)$, and the single-object state transition density is defined by $f_+(\cdot|\cdot,\ell_+)$. The birth space at $k+1$ is $\mathbb{B}_+$. The birth probability associated with the label $\ell_+$ is $r_{B_+}(\ell_+)$, and its related spatial distribution is $p_{B,+}(x_+,\ell_+)$;
\begin{gather}
    \psi_{Z_+}^{(j)} (x_+,\ell_+) =
		\begin{cases}
			\dfrac{P_D(x_+,\ell_+)g(z_j|x_+,\ell_+)}{\kappa(z_j)}, &  j \in \{1:|Z_+|\}, \\
			1-P_D(x_+,\ell_+), & \text{otherwise;} \notag
		\end{cases}
\end{gather} 
$P_D(\cdot,\ell_+)$ represents the detection probability corresponding to the label $\ell_+$. The spatial clutter intensity, distributed as Poisson, is denoted by $\kappa(\cdot)$. Lastly, the likelihood of producing measurement $z$ from the single-object state $x_+$ associated with label $\ell_+$ is given by $g(z|x_+,\ell_+)$.

\section{The Proposed Method} 
\subsection{LMB Filtering Recursion}
The LMB filter approximates the GLMB filter by matching the first moment (PHD); hence, the LMB filter is often referred to as the PHD filter for multi-object trajectory estimation. 
Suppose the LMB filtering density at the current time $k$ is $\boldsymbol{\pi} = \{ r(\ell),p(\cdot,\ell) \}_{\ell \in \mathbb{L}} = \{ (w(I),p): I \in \mathcal{F}(\mathbb{L}) \}$. Since the LMB filter is closed under the prediction but not the update step, given the LMB birth model, a joint prediction and update step with measurement $Z_+$ for the LMB density yields the GLMB density, 
\begin{align}
    \boldsymbol{\pi}^g_+(\cdot|Z_+) = \bigg\{ \big(w^{(\theta_+)}_{Z_+}(I_+), p_{Z_+}^{(\theta_+)}\big): (\theta_+,I_+) \in \Theta_+ \times \mathcal{F}(\mathbb{L}_+) \bigg\} \notag
\end{align}
which is then approximated as an LMB density:
\begin{align}
    \boldsymbol{\pi}_+(\cdot|Z_+) &= \bigg\{ \big(r_{Z_+}(\ell),p_{Z_+}(\cdot,\ell) \big) \bigg\}_{\ell \in \mathbb{L}_+} \\
    &= \bigg\{ (w(I_+),p_{Z_+}): I_+ \in \mathcal{F}(\mathbb{L}_+) \bigg\}
\end{align}
with the same first moment as $\boldsymbol{\pi}^g_+(\cdot|Z_+)$ by choosing
\begin{align}
    r_{Z_+}(\ell) &= \sum_{(\theta_+,I_+) \in \Theta_+ \times \mathcal{F}(\mathbb{L}_+)} 1_{I_+}(\ell) w^{(\theta_+)}_{Z_+}(I_+), \label{eq:r_glmb2lmb} \\
    p_{Z_+}(x,\ell) &\propto \sum_{\theta_+ \in \Theta_+} p^{(\theta_+)}_{Z_+}(x,\ell) \sum_{I_+ \subseteq \mathbb{L}_+} 1_{I_+}(\ell) w^{(\theta_+)}_{Z_+}(I_+). \label{eq:p_glmb2lmb}
\end{align}

\subsection{Smooth Trajectory Estimator for LMB Filter}
Given the LMB density, extracting multi-object estimates using optimal methods like the joint or marginal multi-object estimators is intractable~\cite{mahler2014advances}. Typically, a less-than-optimal estimator is used by determining the maximum a posterior (MAP) cardinality estimate from the cardinality distribution, denoted as $\rho$, given by
\begin{align}
    \rho(n) = \delta_{n}(|I_+|) \sum_{I_+ \in \mathcal{F}(\mathbb{L}_+) } w(I_+). 
\end{align}
From the cardinality distribution $\rho$, we can compute the estimated cardinality $\widehat{N}$, given by
\begin{align}
    \widehat{N} = \argmax_{n} \rho(n).
\end{align}
By sorting the existence probability vector $\{r_{Z_+}(\ell)\}_{\ell \in \mathbb{L}_+}$ in descending order, we can choose the foremost $\widehat{N}$ labels based on the highest existence probability within $\boldsymbol{\pi}_+(\cdot|Z_+)$ and let $r_{\min}(\widehat{N})$ be the smallest existence probability from these top $\widehat{N}$ labels.  Therefore, the standard MAP  multi-object state estimates $\widehat{\mathbf{X}}_+$ at time $k+1$ is computed as follows:
\begin{align}
    \widehat{\mathbf{X}}_+ &= \big\{ (x,\ell): r_{Z_+}^{(\ell)} \geq r_{\min}(\widehat{N}), x = \int y p_{Z_+}(y,\ell)dy \big\}. 
\end{align}
Notably, in the standard LMB filter, the association map $\theta_+$ was marginalised over the association space $\Theta_+$ and discarded (see \eqref{eq:r_glmb2lmb} and \eqref{eq:p_glmb2lmb}). However, this association map is crucial to constructing the smooth trajectory estimator for estimating trajectory using all measurements and should not be discarded. In this work, we propose to store the \textit{best weighted} association map of each  label $\ell \in \mathbb{L}_+$, i.e.,
\begin{align}
    \theta_+^{(*,\ell)} = \argmax_{\theta_+ \in \Theta_+} 1_{\mathcal{F}(\mathbb{L}_+)}(I_+) 1_{I_+}(\ell) w^{(\theta_+)}_{Z_+}(I_+),
\end{align}
and the set of the \textit{best} association map of all labels in $\mathbb{L}_+$ is denoted as $\Theta_+^{(*)} = \{ \theta_+^{(*,\ell)}: \ell \in \mathbb{L}_+ \}$. 
The association history of each  label $\ell \in \mathbb{L}_+$ is recursively stored from the birth time $s(\ell)$ to the current time $k+1$, i.e., 
\begin{align}
    \xi^{(*,\ell)}_+ =  \big(\xi^{(*,\ell)},\theta_+^{(*,\ell)}). 
\end{align}
Let $\Xi^{(*)}_{1:k+1} = \{ \xi^{(*,\ell)}: \ell \in \mathbb{L}_{1:k+1} \}$ be the set of the best association history of all labels in $\mathbb{L}_{1:k+1}$.  As a result, by utilising the entire association history $\xi^{(*,\ell)}_{s(\ell):k+1}$ of the trajectory with label $\ell$, we can efficiently estimate the entire history of this trajectory. The detailed algorithm is provided in Algorithm~\ref{algo:STE_LMB}  where $\boldsymbol{\mathcal{X}}_{1:k+1}$ is denoted as the recursive multi-object trajectory estimates from time $1$ to $k+1$. In particular, to estimate the entire history of each label $\ell$, we apply forward filtering via any applicable single-object-tracking filter (e.g., Kalman Filter, Unscented Kalman Filter, or Particle Filter) in lines $7-13$ and then backward smoothing via Rauch–Tung–Striebel (RTS) smoother~\cite{rauch1965maximum} in lines $14-15$.

\begin{algorithm}[!tb]
    \DontPrintSemicolon
    \small
    \KwInput{$\boldsymbol{\mathcal{X}}_{1:k}$, $\Xi^{(*)}_{1:k}$, $\widehat{\mathbf{X}}_{k+1}$, $\Theta^{(*)}_{k+1}$, $Z_{1:k+1}$}
    \KwOutput{$\boldsymbol{\mathcal{X}}_{1:k+1}$}

    \vspace{0.2cm}

    $\boldsymbol{\mathcal{X}}_{1:k+1} \gets \emptyset$ \hfill \tcp{initialisation} 
    
    $\Xi^{(*)}_{1:k+1} \gets \Xi^{(*)}_{1:k} \cup \Theta^{(*)}_{k+1}  $ \hfill \tcp{update history} 

    $\widehat{\mathbf{X}}_{1:k+1} \gets \boldsymbol{\mathcal{X}}_{1:k} \cup \widehat{\mathbf{X}}_{k+1} $ \hfill \tcp{update estimates} 

    $\widehat{\mathbf{L}}_{1:k+1} \gets \mathcal{L}(\widehat{\mathbf{X}}_{1:k+1}) $  \hfill \tcp{extract labels}

    \ForEach{trajectory with $\ell \in \widehat{\mathbf{L}}_{1:k+1}$}{%
        Compute $s(\ell)$ via \eqref{eq:label_birth_time} and $t(\ell)$  via \eqref{eq:label_end_time}

        \vspace{0.2cm}
        \tcc{\phantom{AAAAAAAA} Forward filtering}
        \For{ $i=s(\ell)$ \KwTo $t(\ell)$ \KwBy  $1$ }{ 

            \eIf{$i = s(\ell) $}{
                $x_{i}^{(\ell)} \gets \int y p_B(y,\ell)dy$
            }
            {
                $x_i^{(\ell)} \gets \textrm{SingleObjectPredict}(x_{i-1}^{(\ell)}) $ 
            }

            \If{$\theta^{(*,\ell)}_i >0$}{
                $x_i^{(\ell)} \gets \textrm{SingleObjectUpdate}(x_{i-1}^{(\ell)},Z_{i}(\theta^{(*,\ell)}_i)) $ 
            }
        }
        
        \vspace{0.18cm}
        \tcc{\phantom{AAAAAAA} Backward smoothing}
        \For{ $i=t(\ell)-1$ \KwTo $s(\ell)$  \KwBy  $-1$ }{
            $x_i^{(\ell)} \gets \textrm{SingleObjectSmoothing}(x_{i+1}^{(\ell)}) $ 
        }
        $\mathbf{x}^{(\ell)}_{s(\ell):t(\ell)} \gets \bigg[(x^{(\ell)}_{s(\ell)},\ell),\dots,(x^{(\ell)}_{t(\ell)},\ell)\bigg]$

        $\boldsymbol{\mathcal{X}}_{1:k+1} \gets \boldsymbol{\mathcal{X}}_{1:k+1} \cup \mathbf{x}^{(\ell)}_{s(\ell):t(\ell)} $
      
    }
    \KwRet{$\boldsymbol{\mathcal{X}}_{1:k+1}$}

\caption{Smooth Trajectory Estimator for LMB}  
\label{algo:STE_LMB}
\end{algorithm}

\section{Experiments}
\subsection{Scenario 1 - Linear }
We examine a straightforward linear setting from ~\cite{vo2014glmb}, where we track a variable number of mobile objects (as many as $12$) with different birth and death instances. This occurs within a 2D region measuring $[-1000,1000]$ m by $[-1000,1000]$ m. The duration of this scenario stands at $100$~s.

\begin{figure}[!tb] 
    \centering
    \includegraphics[width=0.48\textwidth]{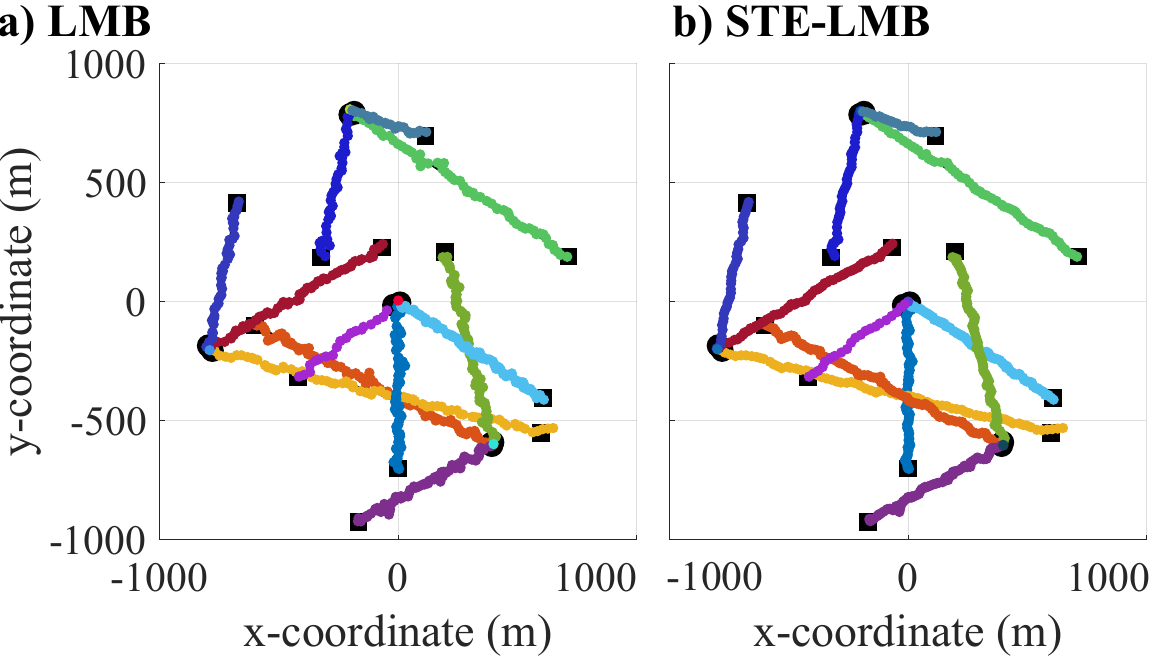}
    \vspace{-0.3cm}
    \caption{\textbf{Scenario 1 (Linear)}: Truth vs Estimates using a) LMB filter, and b) STE-LMB filter.}
    \label{fig:s1_est_vs_truth}
\end{figure}

\begin{figure}[!tb] 
    \centering
    \includegraphics[width=0.3\textwidth]{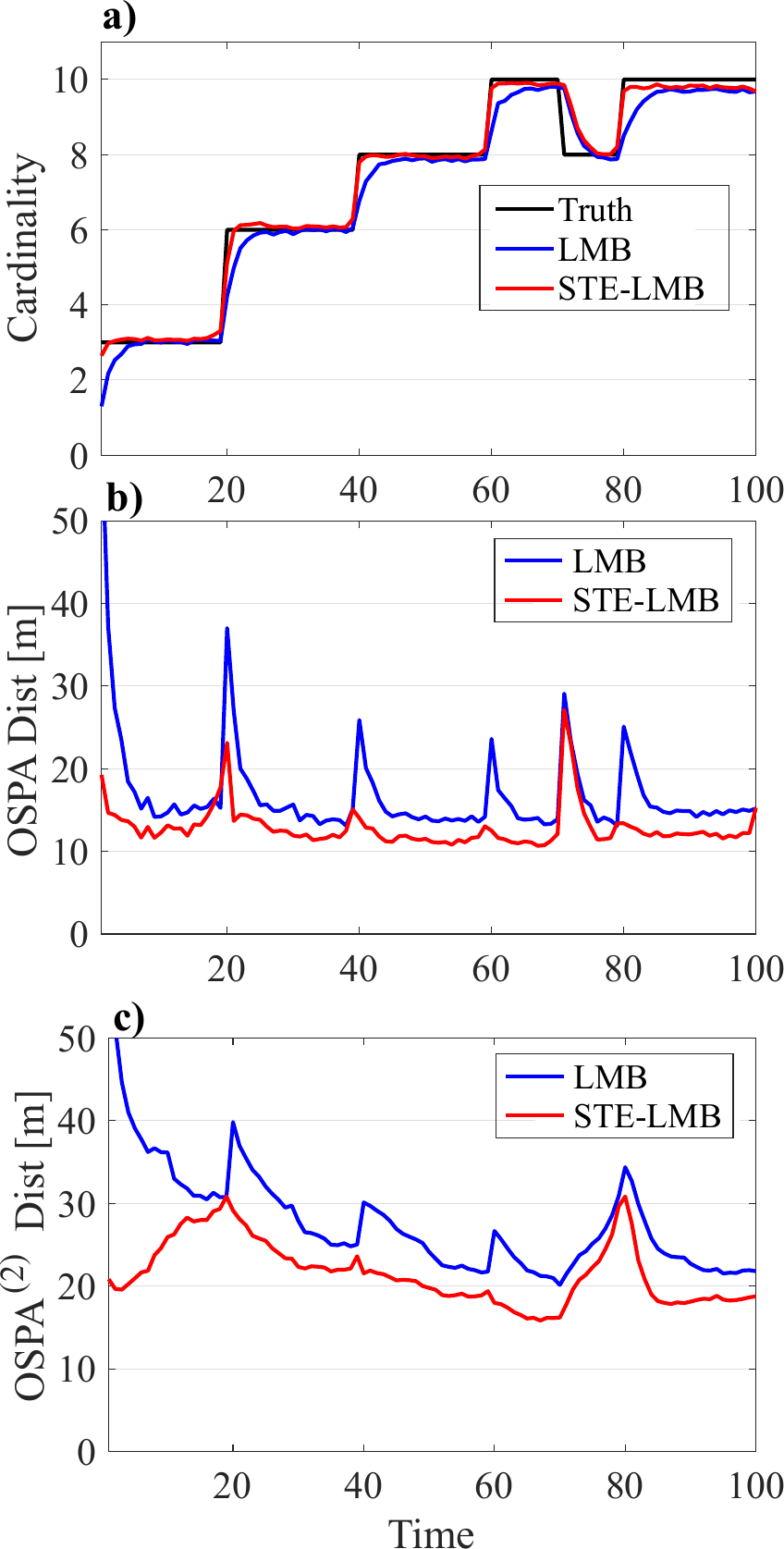}
    \vspace{-0.3cm}
    \caption{\textbf{Scenario 1 (Linear)} - performance comparison results averaged over $100$ Monte Carlo trials: a) Carnality estimation, b) OSPA distance, and c) OSPA\textsuperscript{(2)} distance.  }
    \label{fig:s1_card_ospa_ospa2}
\end{figure}

\textbf{Object dynamic model:} Employing a linear constant velocity model for object dynamics in a 2D setting, each object's state, $x=[p_x,\dot{p}_x,p_y,\dot{p}_y]^T$, encapsulates its kinematic status. The dynamic density is expressed as $f_+(x_+|x) = \mathcal{N}(x_+;Fx,Q)$, where $\mathcal{N}(\cdot;\mu,\Sigma)$ symbolises a Gaussian density. Given that $F=[1,\triangle;0,\triangle] \otimes I_2$ and $Q=\sigma^2_x[\triangle^3/3,\triangle^2/2;\triangle^2/2,\triangle] \otimes I_2$, with $I_2$ being a 2x2 identity matrix, $\otimes$ denoting the Kronecker tensor product, a sampling interval of $\triangle=1$~s, and $\sigma_x=5$~m/s\textsuperscript{2}. Each object $x$ has a 0.99 survival probability, represented as $P_S = 0.99$. For every time increment, the birth density is represented by the LMB density $\boldsymbol{\pi}_B$, comprising $\{r_B(\ell_j),p_B(\cdot,\ell_j) \}_{j=1}^4$. Here, $r_B(\ell_j) = 0.05,~\forall j=1\dots4$ and $p_B(x,\ell_j) = \mathcal{N}(x;\mu_B^{(j)},P_B)$, $\mu_B^{(1)} = [0.1,0,0.1,0]^T$, $\mu_B^{(2)} = 100\cdot[4,0,-6,0]^T$, $\mu_B^{(3)} = 100\cdot[-8,0,-2,0]^T$, $\mu_B^{(4)} = 100\cdot[-2,0,8,0]^T$, $P_B = \mathrm{diag}(10\cdot[1,1,1,1]^T)^2$. 

\textbf{Measurement model:} Each detected object $x$ with a detection probability $P_D = 0.88$ generates a noisy 2D position measurement $z=[z_x,z_y]^T$ with measurement likelihood $g(z|x) = \mathcal{N}(z;Hx,R)$, where $H = [I_2,0_2]$, $0_2$ is the $2\times2$ zero matrix, $R = \sigma^2_r I_2$, $\sigma_r=10$~m. The measurement set $Z$ at each time also contains the clutters  that follow the Poisson model with a uniform clutter density $\kappa(\cdot) = 5.5 \cdot 10^{-7}$ resulting in an average of $66$ clutters per time step. Importantly, due to the linear nature of the dynamic and measurement models, we employ the Kalman filter alongside the conventional RTS smoother for STE-LMB as detailed in Algorithm~\ref{algo:STE_LMB}.

Fig.~\ref{fig:s1_est_vs_truth} depicts the estimated trajectories versus true trajectories using a) LMB and b) STE-LMB.  Comprehensive comparative analysis, averaged across 100 Monte-Carlo (MC) trials between LMB and STE-LMB, is depicted in Fig.~\ref{fig:s1_card_ospa_ospa2}. The results confirm that using the smooth trajectory estimator, we can estimate the multi-object trajectories correctly in terms of cardinality (i.e., the number of time-varying objects), OSPA~\cite{schuhmacher2008consistent} and OSPA\textsuperscript{(2)}~\cite{beard2017ospa2,beard2020solution} errors. 

\subsection{Scenario 2 - Non-Linear }
In the following section, we examine a non-linear setting from~\cite{reuter2014lmb} wherein we track an uncertain, fluctuating count of mobile objects (maximally $10$) that possess different birth and death timings. This takes place within a 2D space spanning $[-2000,2000]$ m by $[0,2000]$ m. The duration of this scenario is $100$ seconds.

\begin{figure}[!tb] 
    \centering
    \includegraphics[width=0.35\textwidth]{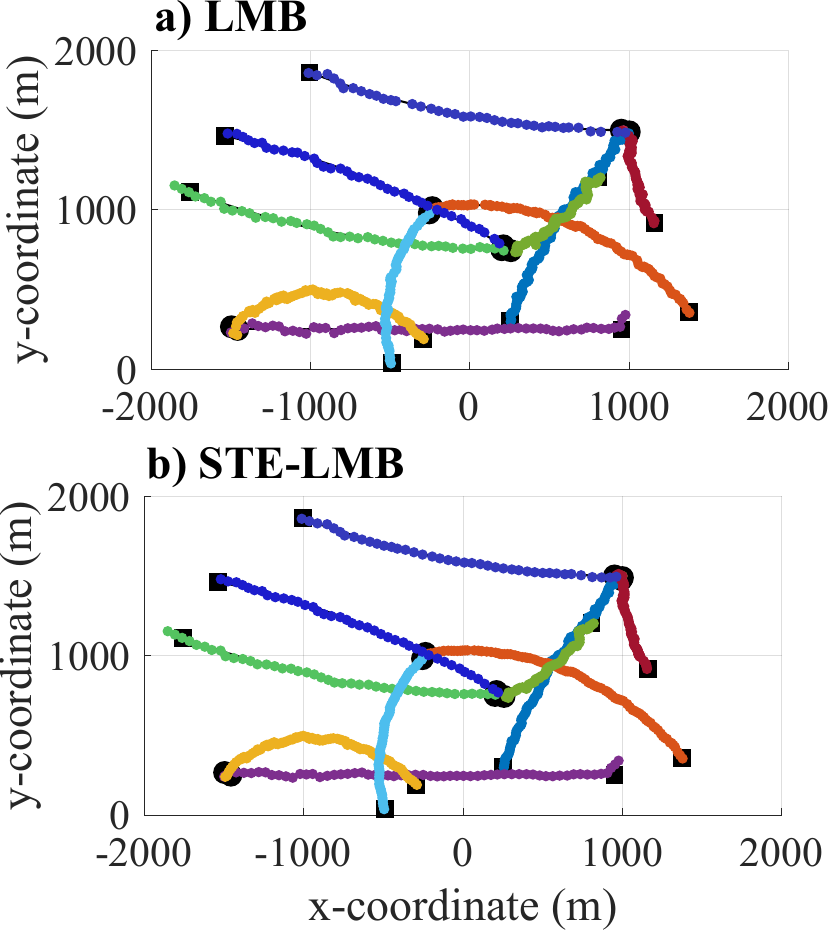}
    \vspace{-0.3cm}
    \caption{\textbf{Scenario 2 (Non-Linear)}: Truth vs Estimates using a) LMB filter, and b) STE-LMB filter.}
    \label{fig:s2_est_vs_truth}
\end{figure}

\begin{figure}[!tb] 
    \centering
    \includegraphics[width=0.3\textwidth]{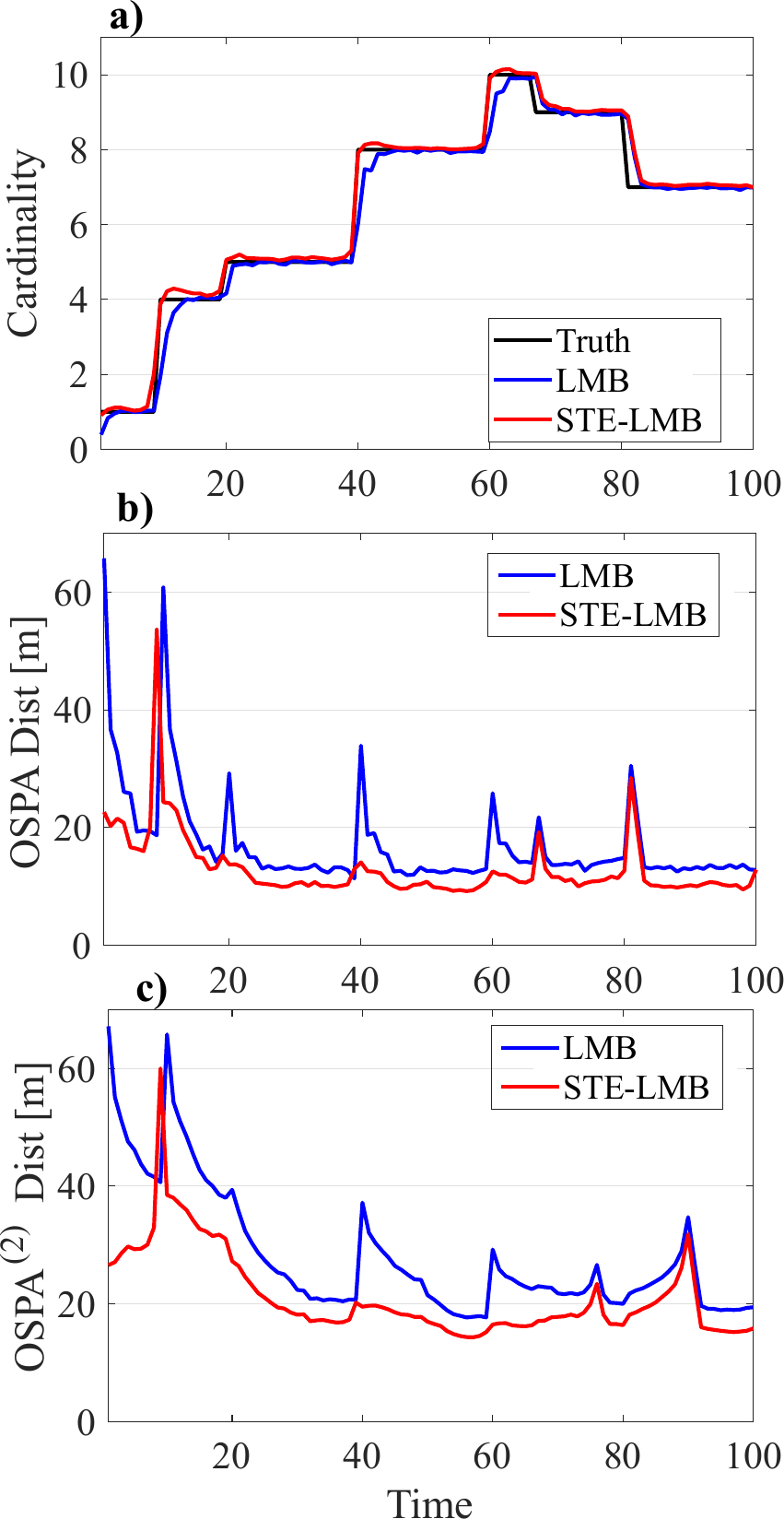}
    \vspace{-0.3cm}
    \caption{\textbf{Scenario 2 (Non-Linear)} - performance comparison results averaged over $100$ Monte Carlo trials: a) Carnality estimation, b) OSPA distance, and c) OSPA\textsuperscript{(2)} distance.  }
    \label{fig:s2_card_ospa_ospa2}
\end{figure}

\begin{figure}[!tb] 
    \centering
    \includegraphics[width=0.37\textwidth]{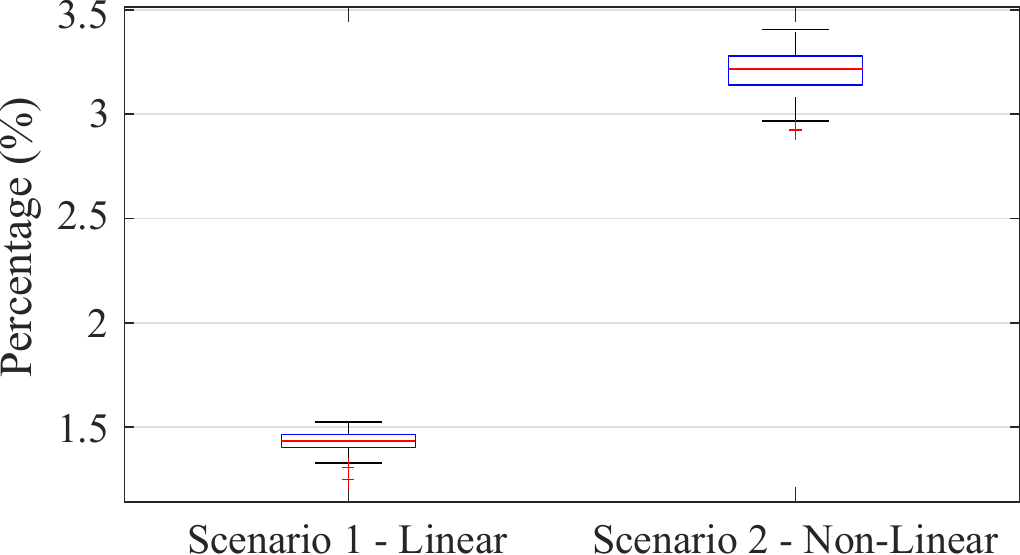}
    \vspace{-0.3cm}
    \caption{Percentage of the smooth trajectory estimator's computational time in relation to the total filtering time averaged over $100$ Monte Carlo trials. }
    \label{fig:compare_time}
\end{figure}

\textbf{Object dynamic model:} Given the non-linear dynamics of the object, we adopt the coordinated turn (CT) model. The single object state is denoted by $x = [\tilde{x}^T, \omega]^T$, wherein $\tilde{x} = [p_x,\dot{p}_x,p_y,\dot{p}_y]^T$ represents the kinematic state and $\omega$ signifies the turning rate. The transition density for the CT model is defined as $f_+(x_+|x) = \mathcal{N}(x_+;m(x),Q)$. Here, $m(x) = \big[[F(\omega)\tilde{x}]^T,\omega\big]^T$. The matrix $Q$ is represented as $\mathrm{diag}([\sigma^2_{\tilde{x}} G G^T,\sigma^2_{\omega}]^T)$, with $\sigma_{\tilde{x}} = 5$~m/s\textsuperscript{2} and $\sigma_{\omega} = \pi/180$~rad/s.
\begin{gather}
        F(\omega) = \begin{bmatrix}
                         1  &  \dfrac{\sin(\Omega)}{\omega} & 0 & -\dfrac{1 - \cos(\Omega)}{\omega} \\
                         0  &  \cos(\Omega)  &  0  & -\sin(\Omega)  \\
                         0  & \dfrac{1-\cos(\Omega)}{\omega}  & 1  &  \dfrac{\sin(\Omega)}{\omega}  \\
                         0  & \sin(\Omega)  &  0  &  \cos(\Omega) \\
                         \end{bmatrix}, \notag 
        G= \begin{bmatrix}
                 \dfrac{\triangle^{2}}{2}  & 0 \\
                 	\triangle  & 0 \\
                 0  &  \dfrac{	\triangle^{2}}{2} \\
                 0  &  	\triangle^{2} \\
                 \end{bmatrix}
\end{gather}    
where $\Omega=\omega \triangle$, $\triangle =1$~s is the sampling interval.  At every interval, each object $x$ progresses to the subsequent time through the dynamic density $f_+(x_+|x)$, having a survival probability of $P_S=0.99$. We adopt an LMB birth model denoted by $\boldsymbol{\pi}_B = \{r_B(\ell_j),p_B(\cdot,\ell_j) \}_{j=1}^4$. Specifically, $r_B(\ell_1)=r_B(\ell_2) = 0.02$, $r_B(\ell_3)=r_B(\ell_4) = 0.03$,  $p_B(x,\ell_j) = \mathcal{N}(x;\mu_B^{(j)},P_B)$, $\mu_B^{(1)} = 250\cdot[-6,0,1,0,0]^T$, $\mu_B^{(2)} = 250\cdot[-1,0,4,0,0]^T$, $\mu_B^{(3)} = 250\cdot[1,0,3,0,0]^T$, $\mu_B^{(4)} = 250\cdot[4,0,6,0,0]^T$, $P_B = \mathrm{diag}([50,50,50,50,\pi/30]^T)^2$. 

\textbf{Measurement model:} For each detected object $x$, a range-and-bearing measurement $z = [r,\theta]^T$ is produced with a detection probability $P_D=0.9$. The measurement likelihood is given by $g(z|x) = \mathcal{N}(z;h(x),R)$ where $h(x) =\big[ \sqrt{p^2_x + p^2_y}, \mathrm{atan2}(p_y,p_x) \big]^T$. The matrix $R$ is represented by $\mathrm{diag}([\sigma^2_r,\sigma^2_\theta])$, with $\sigma_r = 10$~m and $\sigma_\theta = 2\pi/180$~rad. Each measurement $Z$ at any given time step is further disrupted with clutters (false alarms). The clutter follows a Poisson distribution with a uniform clutter density $\kappa(\cdot) = 1.59 \cdot 10^{-4}$, resulting in an average of $15$ clutters per observation. Crucially, due to the non-linear characteristics of both the dynamic and measurement models, we utilise the Unscented Kalman filter and the Unscented RTS smoother for the STE-LMB filter, as detailed in Algorithm \ref{algo:STE_LMB}.

Fig.~\ref{fig:s2_est_vs_truth} illustrates the comparison between estimated trajectories and true trajectories in a specific run, using both the LMB and STE-LMB approaches. The detailed comparison results averaged over 100 MC trials, are depicted in Fig.~\ref{fig:s2_card_ospa_ospa2}. These results validate that employing the smooth trajectory estimator enables accurate estimation of multi-object trajectories in terms of cardinality (i.e., the number of objects that change over time), OSPA, and OSPA\textsuperscript{(2)} errors. Fig.~\ref{fig:compare_time} presents the percentage of the smooth trajectory estimator's computational time in relation to the total filtering time for both two considered scenarios. The additional computational time from the smooth trajectory estimator is insignificant (i.e., less than $3.5\%$) compared to the total filtering time, which demonstrates the efficiency of our method.

\section{Conclusion}

We have devised an innovative and efficient smooth trajectory estimator for the LMB filter. By adopting the intuitive strategy of retaining the optimal association map during the conversion from the GLMB density to the LMB density, our approach offers an efficient smooth trajectory estimator for the LMB filter. This facilitates accurate detection and tracking of a fluctuating number of mobile objects amidst noisy measurements, whilst substantially reducing label switching and track fragmentation. Experimental outcomes highlight our method's superiority over the existing LMB filter, achieved with only a marginal increase in computational time.

\balance

\end{document}